\documentstyle[11pt,aaspp4]{article} \begin{document}
\def\oii{[\ion{O}{2}] $\lambda3727$}
\def\oiii{[\ion{O}{3}] $\lambda5007$}
\input psfig
\lefthead{Axon et al.} \righthead{The emission line region of CSS}

\title{{The morphology of the emission line region \\ of Compact Steep
Spectrum radio sources \footnote{Based on observations with the NASA/ESA
Hubble Space Telescope, obtained at the Space Telescope Science Institute,
which is operated by AURA, Inc., under NASA contract NAS 5-26555 and by
STScI grant GO-3594.01-91A}}}

\author{D. J. Axon\altaffilmark{2,3}} \affil{Space Telescope Science 
Institute, 3700 San
Martin Drive, 21218, Baltimore, MD, USA}

\author{A. Capetti} \affil{Osservatorio Astronomico di Torino \\
Strada Osservatorio 25,
10025 Torino, Italy}

\author{R. Fanti, R. Morganti \altaffilmark{4}} \affil{Istituto di Radioastronomia \\CNR,
via Gobetti 101, I-40129, Bologna, Italy}

\author{A. Robinson} \affil{Division of Physics and Astronomy, \\
 Department of Physical Sciences, 
 University of Hertfordshire,
 College Lane, Hatfield, Herts AL10 9AB,UK} 

\author{R. Spencer} \affil{NRAL, University of Manchester, Jodrell Bank, 
UK}

\altaffiltext{2}{Affiliated to the Astrophysics Division, Space Science
Department, ESA}
\altaffiltext{3}{current address: Division of Physics and Astronomy, \\
 Department of Physical Sciences, 
 University of Hertfordshire,
 College Lane, Hatfield, Herts AL10 9AB,UK}
\altaffiltext{4}{current address: Netherlands Foundation for Research in Astronomy Postbus 2, 7990 AA, Dwingeloo NL}

\authoremail{dja@star.herts.ac.uk,capetti@to.astro.it,rfanti@astbo1.bo.cnr.it,
morganti@nfra.nl, ar@star.herts.ac.uk}

\begin{abstract} 

We present the results of HST narrow band imaging of eleven Compact Steep
Spectrum (CSS) radio sources.  Five of them (3C 48, 3C 147, 3C303.1,
3C 277.1 and 4C 12.50) were observed as part of a dedicated ``pointed''
program of deep line imaging, at the redshifted wavelength of the \oiii\
emission line.  For six additional sources (3C 49, 3C 93.1, 3C 138,
3C 268.3, 3C305.1 and 3C343.1) ``snapshot'' images (\oiii\ or \oii)
were taken from the HST archive.

In all but one of the targets (3C 49) line emission has been detected
and only in the case of 3C 138 is it unresolved at the HST resolution.
Three distinct components are found in the CSS emission line morphologies:
1) compact nuclear emission regions whose size is less than a few kpc:
2) bright emission spatially related to the radio structure; and 3)
faint emission which extends well beyond the radio source.

A large fraction of the line emission (between 30\% and 90\%) originates
within less than $\sim$ 3 kpc from the nucleus as in the case of Seyfert
and extended radio galaxies.

In four out five of the sources for which deep observations are available
the line emission extends well beyond the size of the radio source but
along the radio axis. Only in the case of the largest radio source (3C
277.3) no emission beyond the radio lobes is detected.  Structures of 
similar surface
brightness would have not been seen in the snapshot images. These
emission line structures extend to scales of 10 to 30 kpc and cover
a projected angle, when seen from the nucleus, of  $\sim 30^\circ$ -
110$^\circ$, and indicate that the nuclear illumination is anisotropic.
Photon counting arguments also support this interpretation. In agreement
with the AGN unified scheme, only the CSS with Broad Line Regions show
a strong unresolved continuum source in the HST images.

In six objects the radio emission extends over more than 1\arcsec\
and the HST resolution is such that a detailed comparison can be made
between radio and optical morphologies.  In these cases the line emission
has an elongated structure, linking the nucleus to the radio-lobes,
possibly tracing the path of the invisible radio jets.

Nevertheless the emission line morphologies do not show the bow shocks at
the extremities of the radio lobes one would expect if they are sources
whose expansion is frustrated by a dense external medium.  Our data
favour the alternative model in which CSSs are the young phase of the
large size radio sources.

When ``pointed'' pure continuum images are available, there appears to
be no alignment between radio and continuum emission which contradicts
previous suggestions based on broad-band HST imaging.  We suggest that 
these
broad band images are in most cases heavily contaminated by line emission,
producing a spurious apparent alignment.

\end{abstract}


\section{Introduction}

In the unified scheme for active galactic nuclei (AGN) broad and narrow
line objects are intrinsically the same, but are viewed at different
orientations. The orientation dependence arises either as a result of
preferential obscuration created by a surrounding
torus of dust material, as in Seyfert nuclei, 
in powerful radio
galaxies, as a result of relativistic effects (Antonucci 1993, Urry \&
Padovani 1995). 
It follows that a fundamental question for the unified
scheme is to establish the degree of anisotropy of the radiation field in
 radio loud AGN. The other important question is the balance between the
relativistic beaming and
 radiation cone illumination as a function of radio-power and source
luminosity. 
{\bf In reality the situation can be made more complex 
by secondary effects, such as intrinsic angle dependence of the 
radiation pattern produced by an accretion disk. It is not as yet clear 
if radiation cones exist in the majority of more luminous sources
because these obscuring structures might be evaporated by the radiation 
field (i.e. the absence of quasars type 2).}

The study of the gaseous environment around Active Galactic Nuclei
provides a potent tool for studying both their radiation field and the
mechanical energy carried by the associated radio ejecta. In most of the
radio galaxies studied to date the emission line gas is co--spatial with
the radio emission and therefore contributions to their ionization made by
the turbulent shocks created by the ejecta cannot be distinguished from
photoionization by the nuclear radiation field. However there is a group
of radio sources, known as Compact Steep Spectrum (CSS) sources, where
the radio emission is confined to scales smaller than typical galactic
sizes. Compact steep spectrum (CSS) sources are high luminosity
extragalactic radio sources with steep radio spectra ($\alpha\geq$0.5,
with $S\sim \nu^{-\alpha}$) and small radio angular size ($\leq$2\arcsec,
$\leq$30 kpc) (\cite{fan85}, \cite{spe89}). The CSSs are believed to
represent either a young phase of powerful extragalactic radio sources
(Fanti et al. 1995, Readhead 1995 and references
therein) or radio sources trapped by unusual conditions of their ISM (e.g. 
van Breugel, 1984) which prevents them from growing to normal dimensions.
Hes, Barthel \& Fosbury (1996), Hirst et al.  (1996), Morganti et al. 
(1997) have found that their spectroscopic and polarimetric
characteristics are similar to those of the extended sources of similar
power and redshift. 
{\bf Any emission line gas extending beyond the radio
structure would not be affected by interactions with the radio
ejecta.} In this sense they would be the analog of the Extended Narrow Line
Region (ENLR ) of Seyfert galaxies (Unger et al.  1987). The ENLR of CSS
sources might, therefore, provide a direct probe of the anisotropy of the
radiation field in radio galaxies and QSOs while in the region co-spatial
with the radio ejecta the effects of jet cloud interaction on the
evolution of the radio sources can be investigated. 

Fanti et al.  (1995) have recently investigated if the conditions of the
external medium (warm and hot gas) around CSS sources can keep them small
by confinement.  They conclude that this scenario is unlike, although not
definitely ruled out, and they support the idea of CSS as young phase
of large size radio sources.  Moreover, by studying a small sample of
CSS sources, Gelderman and Whittle (1994, hereafter GW94) have
shown that the profiles of the NLR are broader and complex and suggest
that this is the result of the jet interaction with the external medium.

The limitation of these optical studies is that they are based on
ground-based data where the resolution of the optical observations is not
adequate for these objects.  In this paper we present the results of HST
observations of 11 CSS sources.  The high resolution of the HST allows
us to resolve the morphology of the ionized gas in the region co-spatial
with the radio enabling us to investigate the role of interactions and
their influences on the evolution of these sources.  The plan of the
paper is as follows: in $\S$ 2 we describe the properties of the sample
observed. In $\S$ 3 the observations.  In $\S$ 4 and 5 we present our
results comparing the optical HST images with radio images of similarly
high resolution. The properties of the line emitting regions of CSS
are discussed in $\S$ 6 while the nature and evolution of CSS sources
in the light of these results is discussed in $\S$ 7.  In $\S$ 8 and
9 we analyze the importance of orientation and the alignment effect in
CSS respectively.  Summary and conclusions are given in $\S$ 10.

Throughout this paper we adopt $ H_o = 50$ km s$^{-1}$ Mpc$^{-1}$ and
$q_o = 0.5$.

\section{The Sample and its radio properties}

Our sample comprises 11 objects covering the redshift range 0.12 to 1.12:
 5 of which have been observed with ``pointed'' observations (Table 1)
while the remaining 6 are ``snapshot'' images (Table 2)  taken from the
HST archive.  Of the 11 objects, 4 have permitted broad lines in
 their spectra and are generally classified as Quasars($Q$) while the
remaining 7 only have only narrow lines and are classified as radio
galaxies($G$)

The radio sources of the sample are all very powerful, with radio
luminosities $\geq 10^{26}$ W/Hz at 2.7 GHz.  Their radio data are
summarized in Table 3.  In seven sources, out of eleven, the radio core
has been detected.  The core luminosities are similar to those of radio
galaxies and radio quasars of larger linear size.  The parameter $P_{cn}$
(which represents the ratio of core to extended flux, normalized to the
median value of the object class) is an orientation indicator (Capetti
et al., 1995a).  
The range of $P_{cn}$ values of our sources indicates that the sample is 
not preferentially biased in orientation compared to the population of CSS
as a whole.

The radio morphology is double in 8 out of 11 objects.  We suspect
that also 4C~12.50 might be a double source in which one of the lobes has
been missed (Stanghellini et al. 1997). 3C 48 and 3C 93.1 have rather
peculiar radio structures. In general there are significant asymmetries
between the two lobes, both in flux and arm ratio (ratio of distances
from the lobe edges to the core) and often the lobe closer to the core
is also the brightest one (Fanti et al., 1990, Sanghera et al., 1995).

In Table 3 we have also included in column 10 and 11 the \oiii\ (or \oii)
line fluxes found in the literature.

\section{The HST Observations}

\subsection{Pointed observations}

Narrow band images for 3C 48, 3C147, 3C 277.1, 3C303.1, and 4C12.50, were
obtained in a dedicated program of line imaging of CSS radio galaxies.
The observations were taken  using the Linear Ramp Filters (LRF)
of the Wide Field and Planetary Camera 2 (WFPC2) on board the Hubble
Space Telescope.  With the LRF each CCD pixel is mapped to a unique
central wavelength with a FWHM bandwidth of $\sim$ 1.3 \% of the central
wavelength, which allows the production of narrow band images at any given
wavelength over a field of view of $\sim$ 13\arcsec.  The redshifted
wavelength of the [O III]$\lambda$5007 emission line was selected for
each of the targets (see Table 1 for the log of the observations).

A continuum image was obtained for each target using the LRF centered in
rest frame wavelength range 5400 - 5500 \AA.  The LRF were preferred
to more efficient, broader filters since they allow us to isolate a
region of continuum emission completely free of line emission. Therefore,
although they produce images of relatively lower signal to noise (with a
typical surface brightness limit of $1 - 3 \cdot 10^{-17}$ erg s$^{-1}$
cm$^{-2}$ arcsec$^{-2}$), the genuine continuum structure of the targets
can be effectively explored and compared to the line and radio emission
structure. At these wavelengths the Point Spread Function of HST has a
FWHM of $\sim 0\farcs06$.

In all cases the selected wavelength corresponds to a location in one of
the three Wide Field (WF) CCD chip, where the pixel size is 0\farcs1,
except for 3C277.1 whose on-band image falls into the Planetary Camera
(PC) which has a pixel size of 0\farcs0455.

Three on-band and one off-band images with an exposure time ranging
between 1100 and 1330 s each were taken.  The data were processed through
the PODPS (Post Observation Data Processing System) pipeline for bias
removal and flat fielding (\cite{bur95}). Individual exposures in each
filter were combined to remove cosmic rays events. In the off-band
images cosmic rays have been individually identified and removed by
taking averaged values from neighbouring pixels.

The line and continuum images were aligned by registering
point sources present in both fields of view, except for 3C 48 where
the central point source was used.

The off-band images were scaled to reproduce the continuum contamination
in the on-band images by taking into account the different exposure time
and filter efficiency as derived from the internal WFPC2 calibration
which is accurate to within 5 \%. The 3$\sigma$ sensitivity of the
resulting LRF images is typically  $1.5\cdot 10^{-17}$ erg s$^{-1}$
cm$^{-2}$ for point sources.

\subsection{Snapshot observations}

HST narrow band images for an additional 6 CSS are available in the
HST archive, obtained as part of a ``snapshot'' (two exposures of 300
s each) program of observations of the 3C sample of radiogalaxies (see
table 2 for the observations log). Furthermore, snapshot observations
duplicate our much longer pointed exposures of 3C277.1 and 3C303.1;
we can use these to quantify the relative depth of the two surveys.

The LRF filter were used and centered on the [O III]$\lambda$5007 emission
line for target with z $<$ 0.5, and on the [O II]$\lambda$3727 line for
z $>$ 0.5.  3C 138 was observed through the F656N narrow filter which
covers the redshifted [O III] emission.  As for the pointed observations
the selected wavelength always correspond to locations in the WF chips 
(with
the exception of 3C 138 which was imaged by the PC).

Snapshot broad band images are also available for all targets (one
or two exposures of 140 or 300 s each). The F702W wide filter, which
covers the spectral range 6000 - 8200 \AA \  was used. In all cases the
emission line imaged with the LRF is included in the passband of the
F702W filter. The targets were all located in the PC.

The snapshot data were reduced  using the same procedure as for the
pointed observations.  The 3$\sigma$ sensitivity of the snapshot LRF
images is typically $8\cdot 10^{-17}$ erg s$^{-1}$ cm$^{-2}$ for a 
point source. In general
the snapshot data are read-noise limited and there is therefore a very
large gain in signal to noise in the pointed observations, which are
photon noise limited.

\section{Results}

In all but one of the targets (3C 49) line emission has been detected
and only in the case of 3C 138 it is unresolved at the HST resolution.
A large variety of emission line components are found, compact nuclear
emission regions whose size is less than a few kpc, bright emission
spatially related to the radio structure and faint emission which extends
beyond the radio source.

In Figures 1 through 7 we show the HST and contour radio images of each
source for which we detected extended line emission.  For each target
observed with the pointed observations the top right panel presents the
on-band image and the top left panel shows the off-band image, scaled
to reproduce the continuum contamination in the on-band image which is,
in all cases, negligible.  In order to emphasize low surface brightness
line features a higher contrast line image is presented in the bottom
left panel.  A radio contour map is shown in the bottom right panel.
Given the typical 1\arcsec\ uncertainty in the relative astrometry of
the HST and radio images no attempt has been made to directly overlay
radio and optical images, although in several cases the alignment is
straightforward because there is both a radio core and an optical nucleus.

For the snapshot observations the left and central panels show the F702W
and LRF images respectively. Since the broad band F702W filter includes
the line emission observed with the LRF, and other emission lines, only
upper limits to the continuum contamination can obtained. This prevents
us from obtaining accurate photometry for these objects. However, in all
cases, the on-band image is dominated by line emission.  Conversely,
significant line contamination occurs in the broad band filter images
which explains the similarity observed in many cases between the narrow
and broad band image.

Table 4 gives the total and core \oiii\ fluxes (estimated within a
circular aperture 0\farcs1 \ - 0\farcs3\ in diameter) and the nuclear
continuum flux for the objects with pointed observations.  There is a
good agreement between these values and those found in the literature.

\section{Notes on the individual sources}

\subsection{Pointed Observations}

\subsubsection{3C 48}

Both the on and off-band images of 3C 48 are dominated by a bright central 
nuclear source. Several compact knots {\bf form chains of line emission}
extending to a distance of 6\arcsec North of the nucleus.  These 
features correspond to the structures originally seen by Stockton \& 
MacKenty (1987). The presence of the central saturated point source does 
not allow us to investigate the structure of the central 0\farcs5.  
However, the comparison with the continuum image indicates that significant 
line emission is confined to this region.  In fact, the extended emission 
only accounts for $\sim$ 10 - 15 \% of the [O III] flux measured from 
previous ground-based spectra(GW94).  The GW94 data also show that [O III] 
line profile is very broad and flat-topped.  Our unpublished data at higher 
spectral resolution show that this a consequence of the presence of two 
velocity systems indicating some kind of dynamical interaction, presumably 
with the radio jet.

The best radio images are the VLBI data of Wilkinson et al.  (1991). Two 
knots of high brightness dominate the radio emission in 3C 48.  They are 
separated by $\sim$ 0\farcs05 with a North-South orientation. The southern 
one has a flat spectrum and therefore is identified with the radio core. A 
highly asymmetrical lower brightness emission region extends toward the 
North-East for a total length of $\approx$ 0\farcs8 and is bounded by a 
sharp edge. The small scale radio structure shows a rapid expansion of the 
radio source at $\sim$ 0\farcs1 from the core.  The absence of diffuse, 
more extended emission in lower resolution images (van Breugel et al., 
1992) supports the VLBI result that the the radio emission is confined to 
$\lesssim 1\arcsec$.

\subsubsection{3C 147}

The emission line morphology of 3C 147 is also dominated by an unresolved
central source, embedded in a more diffuse region, elongated along
PA +30$^\circ$ and which extends over $\sim 1\farcs2$, approximately
symmetric with respect to the center of the continuum emission.  Along the
same axis, $\sim$ 1\farcs8 South from the nucleus there is a
faint arc-like structure.  Note that the two compact knots at the eastern
edge of this arc are continuum features (see Fig.  8).  The emission line
arc extends over $\sim$ 1\farcs3 and spans  PA $\sim$ 185$^\circ$
to PA $\sim$ 230$^\circ$.  
Another galaxy, possibly a companion,
is located 3\farcs9 away at PA 190$^\circ$.

The double radio structure (van Breugel et al., 1984) is oriented
along PA $\sim$ +30$^\circ$ and extends over $\sim$ 1\arcsec\ and
it is therefore well aligned and essentially cospatial with the line
emission. The southern component envelopes a very bright jet.  The radio
core is well separated from the surrounding bright 3C 147 steep spectrum
emission at high resolution only (Alef et al., 1991).

\subsubsection{3C 277.1}

The line emission of 3C 277.1 extends over 1\farcs5 in a NW - SE
direction.  The overall structure forms a double shell-like morphology,
with the NW lobe brighter and with a better defined morphology.
The central unresolved source contains $\sim 30 \%$ of the total line
emission.

The radio source has a triple structure of which the central component is
the radio core, characterized by a flat spectrum
(Sanghera et al. 1995, Akujor et al. 1995).  
The two lobes, separated by 1\farcs1 and
0\farcs4 respectively from the core, are oriented along PA -50$^\circ$.
Registering the radio core on the peak of the line emission, we find
that the northern lobe extends beyond the emission line gas,
while the southern one is still embedded within it.

\subsubsection{3C 303.1}

The \oiii\ emission of 3C 303.1 has a striking S-shaped morphology,
reminiscent of the NLR structure of the Seyfert galaxy Mrk 3 (Capetti
et al. 1995b). Its brightest inner structure is oriented along PA $\sim$
-40$^\circ$ and it then swings in a NS direction, extending over $\sim$
3\arcsec.  The central and brightest blob is coincident with the continuum
peak but is clearly dominated by line emission.  An arc-like structure,
perpendicular to the overall orientation of the line emission is located
toward the South at a distance of $\sim$ 1\farcs7 from the nucleus. The
galaxy starlight is highly elongated along PA $\sim$ 0$^\circ$.

The radio emission is an asymmetric double (Fanti et al., 1985),
with the SE lobe brighter than the NW lobe.  It is oriented along PA
-47$^\circ$  and extends over 1\farcs8. No radio core has been detected
yet, so the registration of the radio and optical images is somewhat
uncertain. However, both radio lobes are likely to be located outside
the region \oiii\ emission, due to the change in the optical orientation
for radii larger than 0\farcs5. The arc-like line emission structure to
the south has  no corresponding radio counterpart .

\subsubsection{4C 12.50 (1345+125)}

The optical identification of 4C 12.50 is controversial.  Ground based
imaging by Gilmore \& Shaw (1986) revealed a complex optical morphology,
with two nuclei separated by $\sim$ 1\farcs8\ embedded in a distorted
common envelope. They associated the radio source with the eastern optical
component since this is closer (0\farcs4) to the radio position than the
western nucleus (which is offset by 1\farcs0).  However, new astrometry by
Stanghellini et al. (1993) leads to the opposite result. We measure the
position of the optical nuclei in our HST images. They are located at RA=
$13^h ~ 47^m ~ 33^s.49$, DEC= 12$^o$ ~ 17$^{\prime}$ ~ 23\farcs40\ and
RA= $13^h ~ 47^m ~ 33^s.36$, DEC= 12$^o$ ~ 17$^{\prime}$ ~ 23\farcs85\
(J2000) and the offsets from the radio source (RA= $13^h ~ 47^m ~
33^s.31$, DEC= 12$^o$ ~ 17$^{\prime}$ ~ 23\farcs99) are 2\farcs4\ and
0\farcs6\ for the East and West components respectively. Even considering
the accuracy of the HST absolute astrometry, $\sim$ 1\arcsec, it appears
that the most likely identification of 4C 12.50 is the Western nucleus
and it will adopted in this paper.

The central component of the line emission is very compact and shows a
faint elongation towards the West.  At 1\farcs2 North there is an arc-like
structure, which extends $\sim$ 1\arcsec \ perpendicular to the
direction to the nucleus.  Fainter diffuse emission is found 2\arcsec \
North of the nucleus.

The radio emission (Stanghellini et al., 1997) is confined within $\sim$
0\farcs1 (300 pc).  It shows a distorted, triple morphology oriented
approximately along PA -20$^\circ$.  The compactness of 4C12.50 is
confirmed by the VLA observations of Crawford et al. (1996) in which it
appears unresolved.

\subsection{Snap-shot Observations}

\subsubsection{3C 49}

3C 49 is the weakest source in our sample and there is only marginal
evidence for line emission in the narrow band image which is not
presented. The broad band
image (de Vries et al.  1997) shows a central compact component and a
faint elongation, less than 1\arcsec\ in size, along PA $\sim - 55^\circ$.

The radio source associated with 3C 49 has a double asymmetric structure
elongated in the EW direction (Fanti et al., 1989), with a weak core
closer to the brighter western  lobe.  The overall size is $\sim$
1 \arcsec.

\subsubsection{3C 93.1}

The central optical source is marginally extended along PA + 60$^\circ$ and
is dominated by line emission.  A faint emission-line feature extends $\sim $
1\arcsec \ from the nucleus along $\sim$ 45$^\circ$.  The radio structure
is both complex and compact ($\sim 0.6$ \arcsec), about a factor of 2 smaller 
than the emission-line region  (Dallacasa et al., 1995).

\subsubsection{3C 138}

3C 138 appears unresolved at the resolution and sensitivity of the
snapshot images which are presented. 
However, the central continuum source accounts for
only 15 \% of the on-band emission, indicating that line emission is
associated with its nuclear regions and it originates in a very compact
region, $\lesssim $0\farcs1 (1.5 kpc).  The radio source has a triple
structure, with a bright radio core and a bright jet embedded in the
NE lobe.  Around the radio core position Cotton et al.  (1997) found
a large Faraday rotation measure which is likely to occur in the compact line
emitting  region.  In contrast, in the northern jet/lobe, 
the rotation measure is virtually zero, consistent with the lack of line 
emission indicating the presence of little warm gas.

\subsubsection{3C 268.3}

The line emission has a jet-like morphology which extends 0\farcs7
NE and 2\farcs0 \ toward the SW from the center of the host galaxy.
In the brightest regions it is oriented along PA -45$^\circ$ while
it bends toward smaller position angles in the fainter extensions on
both sides. The central component is also dominated by line emission.
Another galaxy, possibly a companion,
 is seen 2\arcsec \ SW of 3C 268.3.

The radio emission has an asymmetric double lobed morphology (Fanti et
al., 1985), extended $\sim$ 1\farcs4 along PA -20 $^\circ$, clearly
misaligned with respect to the line emitting gas.  A weak core is
detected, closer to the southern component (Ludke et al. 1998).
Registering the radio core on top of the peak of line emission, shows
that the northern lobe is at the edge of the line emission, while the
southern, due to the difference in optical and radio PA,  is outside
the line emission region.

\subsubsection{3C 305.1}

The on-band image shows an elongated structure, extending over $\sim$
1\farcs5 \ along PA $\sim$ +30$^\circ$, which is also seen in the
F702W image.  The comparison between the narrow and broad band images
indicates that this is dominated by line emission.

The radio emission forms a double lobed structure with a separation
between the two components of  $\sim$ 2\arcsec.  The radio axis is at
PA -10$^\circ$.  The radio core has not been detected, so that a good
registration of the optical and radio image is not possible.  However,
almost certainly both radio lobes are outside the line emitting region.

\subsubsection{3C 343.1}

The broad-band image reveals a linear feature, 0\farcs5 in size,
oriented approximately along the EW axis, which is superposed on more
diffuse emission.  The similarity to the structure seen in the narrow
band image indicates that they are dominated by emission lines.

The radio source has a double structure, with overall size $\sim$ 0.38
\arcsec, and is oriented  EW.  The two lobes are different
in flux and shape, the western one being more luminous and broader.
No radio core has been detected yet, so that the registration of the
optical and radio image is somewhat uncertain.  However it is likely,
that both radio lobes are still embedded in the NLR.

\section{The properties of the emission line regions in CSS}

Ground-based spectroscopic studies of CSS have been carried out by
GW94, Baker et al.  (1996), Hirst et al.  (1996)
and Morganti et al.  (1997). CSS radio sources have optical spectra
which, to first order, are typical of powerful radio galaxies. In all
cases, strong emission lines and high (or medium) ionization spectra
have been found.  It is therefore not surprising that all our objects
are detected in our new \oiii\ images, while only one is undetected in
the \oii\ images.

However, one of the new results of this study is that EELR, extending {\sl
beyond} the radio emission, are commonly found in CSSs. Clearly, unlike
the conventional radio galaxies, this extended gas cannot be excited by
interaction with the radio outflow and is presumably therefore photoionized by the
nuclear radiation field.   Furthermore, for the well resolved objects we see a correlation
between radio and line emission structures. In the following sections
we will elaborate on these results.

\subsection{The inner  emission line regions}

In all on-band images, with the possible exception of 3C 49 and 3C
305.1, a central compact (smaller than 0\farcs3) component is seen.
The comparison with the continuum or the broad band images clearly
indicates that these central sources are dominated by line emission.
A large fraction of the line emission (between 30\% and 90\%) originates
within 0\farcs1 \ - 0\farcs3\ of the nucleus, corresponding to a
linear scale of less than $\sim$ 3 kpc.  The narrow filter passband,
which is  less than 90 \AA \ wide, includes only the forbidden [O
III]$\lambda\lambda$4959,5007 (or \oii) lines.  Any contribution from
permitted broad lines is excluded. We are observing compact Emission Line
Regions, commonly observed in extended radio-galaxies, and whose size
is comparable to the typical extension of the NLR of Seyfert galaxies.

\subsection{The extended line emitting regions}

In four of the pointed sources, 3C 48, 3C 147, 3C 303.1 and 4C12.50, the line emission
extends well beyond the radio structure.  Not surprisingly these are
sources for which pointed observations are available. We also note that
the only source for which we have pointed observations which does not
show line emission extending beyond the radio lobes is 3C 277.1 which is 
the largest amongst these
radio sources. We conclude that line emission, on a scale significantly
larger than the radio emission, is commonly detected when the images
are sufficiently deep.

The structure of this extended emission is intriguing.  In three cases
(3C 147, 3C 303.1 and 4C12.50) it takes the form of an arc-like feature
perpendicular to the radio axis, but displaced far beyond the lobe edge.  In 3C 48
it is quite different being concentrated in arm-like structures which are
nonetheless located essentially along the radio axis.  The gas responsible
for this extended emission is localized in well defined structures. Again,
this is very similar to what is observed in the ENLR of Seyfert galaxies
(e.g. NGC 5252, Tadhunter \& Tsvetanov 1989, and Mrk 573, Capetti et
al. 1996) which are composed of arcs and filaments of gas.  Both in CSS
and Seyferts, these structures are found approximately in the direction of
the radio axis but they are elongated in a direction perpendicular to it.

The origin of the illuminated shell structures is unclear. They might
have formed in a previous phase of nuclear activity as a result of compression
of the ISM by  radio ejecta. The lack of associated radio emission 
requires a long time scale between the different nuclear phases.  Alternatively,
these structures might be intrinsic to the gas distribution of the
galaxy. For example, gaseous  shells might have been formed due
to a merger and in this case we only see those parts which are
illuminated by the nuclear radiation field.

The projected angle covered by these structures, as seen from the nucleus,
varies from $\sim$ 30$^\circ$ in 3C 303.1, to $\sim$ 45$^\circ$ in 4C
12.50 and 3C147, to 110$^\circ$ in 3C 48.  Since the line emission is
tracing the intersection between the gas and the geometrical pattern of
the nuclear radiation, either the radiation field is highly anisotropic
or the gas is located only along the radio axis. We will discuss 
the issue of anisotropy further in Section 7 from the perspective of
``photon counting''.

\subsection{The association between radio and optical emission}

In six sources (3C 147, 3C 268.3,  3C 277.1, 3C 303.1, 3C 305.1 and
3C 343.1) the size of the radio emission is such that we can study the
relationship between radio and optical structures.

In 3C 303.1 and 3C 268.3 the radio emission has a double lobed morphology
and the line emission originate in two symmetrical jet-like structures
which connect the central source with the radio lobes.  The images of 3C
305.1 are clearly of lower quality, but this source appears to share a 
similar
elongated morphology.  These linear structures of the line emission
follow the radio axis, suggesting that they are tracing the path of
the undetected radio jets which are feeding the lobes.  It is likely
that the compression caused by interaction between the jets and the
external medium causes the emission to be highly enhanced along their
path (Taylor et al.  1992).  The very broad (FWHM up to 2000 km/s)
and flat topped line profiles commonly observed in CSS (e.g. GW94)
are also an indication of jet-induced gas acceleration. Interestingly,
the line emission is always slightly mis-aligned with respect to the
radio axis and its structure is not exactly straight but clearly curved.
Although it is possible that the invisible jets are indeed bent, this is
not usually the case in the more extended, double lobed, radio source
(e.g.  Cygnus A) in which the highly supersonic jets are quite linear.
The results obtained for these sources are very reminiscent of what
is observed in Seyfert galaxies in which there is a close association
between radio and line emission and similarly a slight misalignment
between the radio jets and optical emission is observed (Capetti et
al. 1995b). Capetti et al. interpreted this as due to the expansion of
the radio source in a stratified gas distribution and it is likely that
this idea is also applicable to the CSSs.

The situation is more complex for 3C277.1: the SE radio lobe is located at
the edge of the line structure. Conversely the NW lobe is well outside the
shell-like line emitting region.  In 3C343.1 and 3C 147 the smaller size
of the radio structure does not allow us to perform a detailed analysis,
but clearly the radio and the line emission are well aligned and of very
similar angular size.

Overall, we do not find any clear connection between radio structure
asymmetry and line emission asymmetry. For instance, in 3C277.1 the
lobe closest to the core is associated with the  brightest 
region  of  line emission,
while the converse is true for 3C268.3.

\section {The ionization mechanism}

Optical spectra of CSS sources are available only for a handful of 3CR
sources from GW94 and Hirst, Jackson \& Rawlings
(1996) and for those amongst the complete 2--Jy sample studied by Morganti
et al. (1997). A comparison of the emission line luminosities of CSS
sources with those of extended radio sources has been recently carried
out by Hes, Barthel \& Fosbury (1996) and by Morganti et al.  (1997)
for the 2Jy sample.

The log L$_{\rm [OIII]}$ vs log P$_{\rm radio}$ plot, including [O III] 
luminosities from the HST data, is shown in Fig.  8 in which CSS sources 
are marked with filled symbols.  Although, admittedly, the spread in the 
correlation is large, for a given radio power, the CSS have [OIII] 
luminosities comparable to those of the extended sources.  Similarly CSS 
quasars and CSS radiogalaxies are indistinguishable on this plot (cf.  Hes 
et al.  1996, and Morganti et al.  1997).

This overall similarity of the CSS and extended sources indicates that they 
have ,  at least to first order, their emission-line regions have similar 
physical conditions and ionization mechanisms .

Typical line ratios of the narrow line component in CSS (GW94) are:

      $H_{\beta}$/\oiii\ ~~ $= ~0.18 ~\pm 0.02$

      \oii/\oiii\  $\approx 0.3 \pm 0.05$

      $(H_{\alpha}+[N II])/H_{\beta} \approx 7.5 \pm 1.5$

These line ratios are consistent with photoionization by a power-law
continuum from the nucleus (Robinson et al. 1987).  Nevertheless, it is
important to check that the nucleus is actually sufficiently luminous
to provide the required ionizing flux.  We do this by comparing the
ionizing photon luminosity determined from the emission line fluxes
using photon-counting arguments with that inferred by extrapolating
the observed optical continuum of the nucleus.  We first calculate the
rest-frame optical luminosity, $L_{\nu_{F}}$, for each source using the
fluxes listed in Table 4.  Representing the optical--X-ray continuum
by a power-law of spectral index, $\alpha_{ox}$, the ionizing photon
luminosity is given by

\begin{equation}
 {Q_{ext} =\frac{ L_{\nu_{ F}} \left( \frac{\nu _{F}}{\nu _{H}}\right)^
{\alpha_{ox}}} {h\left|\alpha_{ ox}\right|}}
    \label{eq:1}
\end{equation}

where $\nu_{F}$ and $\nu_{H}$ are, respectively, the frequencies
corresponding to the filter central wavelength and the Lyman limit.
We adopt the average value of the optical--X-ray spectral index,
$\alpha_{ox}\approx -1.3$ found by Brinkmann et al.  (1997) for radio
loud AGN. The ionizing photon luminosities determined in this way are
listed in Table 5.

The ionizing photon luminosity necessary to sustain the line emission is
easily calculated from the H$\beta$ luminosity.  In order to estimate
the latter we multiply the measured [OIII] $\lambda$ 5007+4959 fluxes
(Table 4) by the factor $3/4\times$ the typical $H\beta/$\oiii\ ratio
for CSS quoted above, and use the result to calculate the rest frame
luminosity. The minimum ionizing photon luminosity required to produce the
$H\beta$ emission corresponds to the limiting case in which all ionizing
photons are absorbed, that is, the covering factor of the emission line
region is unity. This is given by

\begin{equation}
 {Q_{min} = \frac{L_{H\beta}}{p_{H\beta}\,h\nu_{H\beta}}} 
    \label{eq:2}
\end{equation}

where $p_{H\beta}\approx 0.1$ is the probability that any recombination
will result in the emission of an H$\beta$ photon.

We have calculated minimum ionizing photon luminosities for both the
NLR and the EELR; their ratios to $Q_{ext}$
are listed in Table 5.  If the nuclear ionizing continuum is isotropic
these ratios are equivalent to the covering factors of the respective
emission line regions.

There is a clear difference between the two quasars for which we have
emission line fluxes, and the two radio galaxies.  In the quasars,
$<20$\% of the available ionizing photons need to be absorbed in the
NLR to produce the line emission.  Since this fraction seems reasonable
for the NLR covering factor, we conclude that the photon budget is
consistent with pure photoionization by the nuclear continuum source.
The nuclear continuum sources of the quasars are also powerful enough to
photoionize their extended emission line regions.  The implied covering
factors are relatively high ($\approx 0.3 ~and ~0.4$, respectively) but
inspection of the images (Figs 2 and 3) shows extensive \oiii\ emission
widely distributed around the nucleus in both cases.  Furthermore, our
power-law extrapolation may underestimate the true ionizing luminosity
if the ``big blue bump'', which is ubiquitous in quasars, contributes
significantly to the EUV continuum.

For the radio galaxies, on the other hand, the inferred ionizing
luminosity is barely sufficient to power the NLR emission, with covering
factors $\approx 0.5$ and $\approx 1$, respectively, being required.
The ionizing photon budget for the ELR in 3C\,303.1 appears to be even
more difficult to reconcile with nuclear photoionization, since $Q_{min}$
exceeds $Q_{{ext}}$ by a factor $\approx 5$.  This is entirely 
consistent with what we expect on the basis of unified schemes
for radio-loud AGN. These schemes hold that the central continuum 
source and broad-line
region are surrounded by a dusty molecular torus, with radio galaxies
and quasars being identified as ``edge-on'' and ``pole-on'' sources,
respectively.  If this is correct, the optical continuum observed in the
radio galaxies is unlikely to come directly from the active nucleus and
therefore the calculated values of $Q_{{ext}}$ will not reflect the true
nuclear ionizing photon luminosity.  For this reason, we cannot exclude
AGN photoionization of either the NLR or ELR in the two radio galaxies,
even though the calculated covering factors are implausibly high. These
results can be explained if the nuclear radiation field is anisotropic.

Another possibility is that shock ionization is important.  Recently a 
jet-driven auto-ionizing shock model for the line emission has been 
presented by Bicknell et al.  (1997).  In order for emission lines to 
be observed requires both that the velocities are $\leq 10^3$ km 
s$^{-1}$ and that the external densities ($>$ $10^2$ cm$^{-3}$) are 
high, so that the cooling time of the shocked gas is smaller than the 
dynamical time of the radio source.  Of course this mechanism cannot 
apply in those cases in which the emission lines originate beyond the 
radio emission.

As described in Section 6, in all sources in which the size the radio
emission extends over more than 1\arcsec, and which are therefore
sufficiently extended to be fully resolved by our HST images, the
bow shock structures which in this model are expected to enshroud
the advancing radio-lobes are not observed. This indicates that the
leading shocks are already in a non radiative phase when the radio
source size exceeds a typical scale of $\sim$ 5 - 10 kpc and do not
produce significant line emission when compared to the total source line
luminosity. 

In contrast, the innermost regions of line emission in these CSS can be
powered by fast shocks which are laterally expanding in the regions where
the ISM is denser and with a shorter cooling time. This argument also
applies to the unresolved CSS of our sample.  
In general to establish which mechanism dominates, requires a combination 
of UV line diagnostic ratios  (Allen et
al. 1998) and shock velocities determined from kinematic studies. 
In this context spectroscopy with the HST would be critical in separating 
the NLR from the ENLR in CSSs.

\section{The nature and evolution of CSS}

As we described in the Introduction CSSs may be either young or frustrated
radio sources. While it is clear from our present data that the larger
CSS are currently not trapped by ambient gas the question remains if
their earlier evolution was significantly impeded by the environment or
if the smaller CSS in our sample are still frustrated.  Here we use our
new data and data available in the literature to investigate this issue.

De Young (1993) and Fanti et al.  (1995) have shown that, in order to
confine average power CSSs  for periods in excess to $10^7$ years, an
ambient gas of number density $>$ 1 - 10 cm$^{-3}$ (with corresponding
core radii from 5 to 1 Kpc) is required.  The mass of gas implied by the
``frustration scenario'' is always rather large, $\geq 10^9 M_{\sun}$.
If this gas exists, the problem is to find out the kind of medium
(hot, tepid, cold).  While it seems (O'Dea et al., 1996) that the
(limited existing data exclude a hot ($\geq 10^7$) medium, which would
produce copious X-rays it is still possible that a cooler confining 
medium exists.

We can use the line luminosities to investigate  if the density of 
line emitting gas is
adequate to trap the radio ejecta.  From our images we can estimate 
the size of the emission
line regions as well as the associated [OIII] luminosities.  When no HST
photometry is available we used data from GW94.  The [O III] luminosities
are converted to narrow $H_{\beta}$ luminosities using an average ratio
of 0.18 $\pm $ 0.02. We preferred this average value to that measured for
individual objects in GW94 since they include also an unknown contribution
from the broad $H_{\beta}$ line.  Since the HST and GW94 luminosities
are in reasonable good agreement, this implies that they refer to the
same emitting volume.  When the \oiii\ line is not available, we use the
\oii\ luminosity and convert it to $H_{\beta}$ by a factor 0.6 $\pm$ 0.1.

Assuming a case B (Osterbrock 1989) model the  $H_{\beta}$ luminosity
is given by:
\begin{equation}
    L(H_{\beta})~~\approx ~~1.24 ~ \times 10^{-25} ~ n_e^2 ~~ \Phi~ V 
    \label{eq:3}
\end{equation}

where $n_e$ is the electron density of the line emitting region, V its
volume  and $\Phi$ the filling factor.  We find  0.3 $<$  $n_e \times
\Phi^{1/2}$  $<$ 40.  The larger values for $n_e \times \Phi^{1/2}$
are found for 3C 49 and 3C 138, for which the line flux is produced in
a region much smaller than the radio size. Excluding these two objects,
the upper limit on  $n_e \times \Phi^{1/2}$ is $\approx 4$.

As shown by Fanti et al. (1995), if the external medium is clumpy, it is the
volume averaged density $n_{sm} = n_e \times \Phi$ which is appropriate 
in confining the radio source. The value of  $n_e^2 \times
\Phi$  determined using  equation (\ref{eq:3})
are such that, for $\Phi$ $<$ $10^{-2}$, $n_{sm}$ is too small to
trap the CSS.  Conversely, if we take a  representative gas 
density of $n_e \approx 10^3$, determined directly for a few objects 
using the [SII] doublet (Eracleous and Halpern, 1993), 
we get $\Phi \approx 10^{-5}$ and  $n_{sm} \approx
10^{-2}$, at least two  orders of magnitude less than what needed by the
"frustration scenario".

Finally, we note that a cool ambient medium sufficiently dense to confine the
radio source would also be optically thick to ionizing photons. The
required column density is $\geq 2\times 10^{22}$\, cm$^{-2}$, which
for neutral hydrogen, corresponds to a Lyman limit optical depth $\sim
10^{5}$. Few ionizing photons would escape the confining medium and we
would not, therefore, expect to see line emission extending significantly
beyond the radio source as is the case in 4 of the 5 sources for which
we have obtained pointed observations.

We conclude that our HST data favour the model in which CSSs are the
young phase of the large size radio sources.

\section{Is there any radio-optical continuum alignment effect?}

Recently, de Vries et al. (1997) found a strong alignment effect in broad
band HST images of CSS at all redshifts. The off-band images obtained as
part of the pointed observations allow us to study the genuine continuum
morphology of these sources and therefore to explore the origin of
their findings. In fact the LRF images isolate a spectral region free of
emission lines, 100 \AA \ wide centered in rest frame wavelength range
5400 - 5500 \AA. The contour  maps of these five continuum images are
given in Fig. 8.  In the same figures we also mark the orientation of
the radio axis.  The lower contours correspond to a surface brightness
of $1 - 4 \cdot 10^{-17} {\rm erg s}^{-1} {\rm \AA}^{-1} {\rm cm}^{-2}
{\rm arcsec}^{-2}$; these reference values translate approximatively
into a brightness limit of 20 - 22 mag arcsec$^{-2}$ in the V band.

Three of these sources are identified with QSO's and the host galaxy
is only marginally detected: in 3C 147 the lower brightness regions
are slightly elongated along NS; in 3C 48 the diffuse extended emission
appears to be oriented at PA $\sim$ - 30$^\circ$; the host galaxy of 3C
277.1 is essentially circular.

Conversely, in the radio galaxies 3C303.1 and 4C 12.50 the stellar light
is clearly visible: in the first case it is highly elongated in the NS
direction; the case of 4C 12.50 is more complex due to the presence of
a nearby companion and of a common diffuse halo. In the inner regions
the isophotes are asymmetrical, being more extended, in both galaxies,
toward the side opposite to the companion.

The comparison between the radio and the optical continuum axis is
not straightforward since, in most cases, the optical structure is
essentially circular. In 3C 303.1, however, there is a clear misalignment
($\sim 40^{\circ}$) between radio and optical structure. The optical
continuum structure is not elongated in the direction of the radio axis 
in any of our sample.

A possible explanation for these contrasting results might be the relative
depths of the two set of data; however our longer exposure times ($\sim
1200 s$) with respect to the snapshot observations compensates largely
for the reduced efficiency of the linear ramp filter when compared to
the broad F702W filter.

We conclude that broad band images are in most cases heavily contaminated
by line emission which, as we discuss in the previous sections,
is indeed aligned with the radio structure. This interpretation is
strongly supported by the comparison of the images for two objects which
are in common between our and De Vries sample (3C 277.1 and 3C 303.1)
for which a close radio/optical alignment is seen in the broad band
images. By separating line and continuum contribution it is clear that
only the line emission is aligned with the radio structure.  Similarly,
the elongated optical structure (closely aligned with radio axis)  seen
in the broad band images of 3C 268.3, 3C 305.1 and 3C 343.3 is cospatial
with the emission line region as revealed by the narrow band images,
suggesting that line emission is also dominating in these cases.

\section{Summary and conclusions}

In this paper, we analyzed narrow band HST images of eleven CSS radio
sources.  In all but one of the targets (3C 49) line emission has
been detected and only in the case of 3C 138 is it unresolved at the
HST resolution.  As is generally observed in extended radio galaxies a 
large fraction (between 30\% and 90\%) 
of the line emission originates within a few kpc of the active nucleus.

In six galaxies the radio emission is sufficiently extended that a
comparison can be drawn between radio and optical morphologies.  The line
emission has an elongated jet-like structure aligned with the radio axis.
It is likely that this connection between the radio and optical emission
is produced by the interaction of the jets with the ambient medium, in
a similar way to that observed in Seyfert NLR. This interpretation is
also supported  by the very broad and flat topped line profiles commonly
observed in CSS. Measuring the velocity and the size of the expanding
line emitting region provides a unique tool for estimating dynamical
timescales for these sources.

However, in four out five of the sources for which deep observations are
available the line emission extends well beyond the size of the radio
source, up to a radius of 10 to 30 kpc.  The gas responsible for this
extended emission is aligned with the radio axis and is localized in
well defined elongated structures confined to ''broad cones'' covering
angles which vary between 30$^\circ$ and 110$^\circ$.  Although the
origin of the arc structures is unclear, their spatial distribution
suggests that they are illuminated by an anisotropic radiation field.
This interpretation is supported by ``photon counting'' arguments.
The unresolved continuum sources seen in Broad Line CSSs provide enough
photons to ionize the ENLR.  Conversely, in Narrow Line CSSs there is a
clear photon deficit which indicates that the nucleus is hidden to our
direct view, in agreement with the AGN unified scheme.

In no source do we see a bow shock shaped emission line regions
enshrouding the radio lobes (commonly observed in Seyfert galaxies)
which are expected to form if CSS are trapped by a dense external medium.
The question remains if their earlier evolution was significantly affected
by the environment or if the smaller CSS are still frustrated. However,
the average densities of ionized gas derived from our images are at
least two  orders of magnitude less than those needed by the "frustration
scenario".  Our data therefore support the alternative model in which CSSs
are the young phase of the large size radio sources. Future kinematic
studies of the gas should allow determination of dynamical timescale in
the manner used by Capetti et al. (1998) for Seyfert galaxies.

Pure continuum images are available for five CSSs.  No alignment between
radio and continuum emission is found.  The alignment effect seen in
broad-band HST images is due to line contamination which, unlike the
continuum, is indeed aligned with radio emission.

\acknowledgements We would like to thank the anonymous referee for their 
helpful and prompt comments. 
RM acknowledges the Australian DIST for support under a
International Science \& Technology (IS\&T) major grant. AC acknowledges
financial support from the STScI Visitor Program.
AR thanks the Royal Society for financial support.

\newpage

\figcaption
{Narrow off-band and on-band (left and right top panel) images of 3C
48. The off-band image was scaled to reproduce the continuum contamination
in the on-band image. In order to evidence lower brightness line features
a higher contrast line image is presented in the bottom left panel. On
the bottom right panel we reproduce the Merlin radio image from 
Wilkinson et al. 1991. Given
the small extension of the radio emission it is presented enlarged eight
times with respect to the HST images}

\figcaption
{Same as Fig. 1 for 3C 147. Narrow off-band and on-band
(left and right top panels), high contrast on-band image (bottom left)
and VLA radio image, taken from van Breugel et al. 1992 (bottom right)}

\figcaption
{Same as Fig. 1 for 3C 277.1. Narrow off-band and on-band
(left and right top panels), high contrast on-band image (bottom left)
and VLA radio image, taken from  Akujor et al. 1995 (bottom right)}

\figcaption
{Same as Fig. 1 for 3C 303.1. Narrow off-band and on-band
(left and right top panels), high contrast on-band image (bottom left)
and Merlin radio image, taken from Sanghera et al. 1995 (bottom right)}

\figcaption
{Same as Fig. 1 for 4C 12.50. Narrow off-band and on-band
(left and right top panels), high contrast on-band image (bottom left).
The VLA radio image, taken from  Stanghellini et al. 1997
(bottom right) is enlarged by a
factor of 10 with respect to the HST images. The radio core is component A.}

\figcaption
{F702W broad band image (left panels), narrow on-band image
(central panels) and the VLA radio images, taken from Dallacasa et al. 1995
and Fanti et al., 1985 (right
panels) for 3C 93.1 and 3C 268.3 respectively}

\figcaption
{F702W broad band image (left panels), narrow on-band image
(central panels) and the VLA radio images, taken from van Breugel et al. 1992
and Fanti et al. 1985 (right
panels) for 3C 305.1 and 3C 343.1}

\figcaption
{Continuum images in the rest frame spectral range 5400 -
5500 \AA \ obtained from the pointed observations of 3C 48, 3C 147, 3C 277.3,
3C 303.1 and 4C 12.50. The dashed lines mark the orientation of the radio axis}

\clearpage
\begin{deluxetable}{lccccccc} 
\footnotesize 
\tablecaption{Log of the pointed observations} 
\tablenotetext{}{Radio-galaxies are identified as $G$ in column 2, while 
quasars are marked as $Q$.}
\tablewidth{0pt} 
\tablehead{ \colhead{Name}
	& \colhead{ID} & \colhead{Redshift} & \colhead{Em. Line} & 
\colhead{$\lambda$
	On-Band} & \colhead{Exp. Time (s)} &   \colhead{$\lambda$
	Off-Band} & \colhead{Exp. Time (s)}    }
\startdata 
3C48    & Q & 0.367 & [O III] & 6845 & 3500 & 7500 & 1100 \nl
3C147   & Q & 0.545 & [O III] & 7736 & 3900 & 8500 & 1200 \nl 
3C277.1 & Q & 0.321 & [O III] & 6609 & 3300 & 7100 & 1100 \nl
3C303.1 & G & 0.267 & [O III] & 6344 & 4200 & 6858 & 1300 \nl 
4C12.50 & G & 0.120 & [O III] & 5608 & 3300 & 6100 & 1100 \nl
\enddata 
\end{deluxetable}


\begin{deluxetable}{lcccccc} \footnotesize 
\tablecaption{Log of the snapshot observations} 
\tablehead{ \colhead{Name}
	& \colhead{ID}& \colhead{Redshift} & \colhead{Em. Line} & 
\colhead{$\lambda$
	On-Band} & \colhead{Exp. Time (s)} &   \colhead{F702W Exp. Time
	(s)}    }
\startdata 
3C49       & G & 0.621 & [O II]  & 6041 & 600 & 300\nl 
3C93.1     & G & 0.243 & [O III] & 6224 & 600 & 300\nl 
3C138      & Q & 0.758 & [O II]  & 6556 & 600 & 280\nl 
3C268.3    & G & 0.371 & [O III] & 6865 & 600 & 600\nl
3C277.1    & Q & 0.321 & [O III] & 6634 & 600 & 280\nl 
3C303.1    & G & 0.267 & [O III] & 6344 & 600 & 300\nl
3C305.1    & G & 1.132 & [O II]  & 7946 & 600 & 600\nl 
3C343.1    & G & 0.750 & [O II]  & 6522 & 600 & 300\nl
\enddata 
\end{deluxetable}

\begin{deluxetable}{lccccccccccc}  \footnotesize 
\tablecaption{Radio parameters of the sample}
\tablenotetext{1}{Units of the radio power W Hz$^{-1}$}
\tablenotetext{2}{Units of the line fluxes 10$^{-15}$ erg s$^{-1}$ 
cm$^{-2}$}
\tablenotetext{a}{[OII] from Jackson \& Browne, [OIII] from Gelderman \& 
Whittle}
\tablenotetext{b}{[OII] from Wills et al., [OIII] from Gelderman \& 
Whittle} 
\tablenotetext{}{References: 1) Gelderman \& Whittle 1994; 2) Jackson \& 
Browne 1991;
      3) McCarthy 1988;  4) Wills et al. 1993;  5) McCarthy et al. 1995} 
\tablewidth{500pt} 
\tablehead{ \colhead{Name}
	& \colhead{$z$} & \colhead{ID} & \colhead{Ang. Size(\arcsec)} & \colhead{Lin. 
Size (kpc)} & \colhead{log ${\rm P}_{2.7GHz}^1$} &   \colhead{log Pc$^1$} & 
\colhead{Sc/S$_{ext}$}  & \colhead{P$_{cn}$}  & \colhead{F$_{\rm 
[OII]}^2$}  & \colhead{F$_{\rm [OIII]}^2$}   &  \colhead{R
ef.}}
\startdata 
3C  48   & 0.37 & Q & 0.8  &  4.9  &  27.76  &   25.78  &  .018   &  .45 & 
13.0 & -- & 1 \nl
3C  49   & 0.62 & G & 1.0  &  7.7  &  27.46  &   24.95  &  .004   & 1.6 &  
3.5 & -- & 3,5 \nl
3C  93.1 & 0.24 & G & 0.6  &  2.9  &  26.51  &  $<$25.0 & $<$.3   &   & 
6.0 & -- & 1   \nl
3C 138   & 0.76 & Q & 0.7  &  5.7  &  28.18  &   26.66  &  .04    & 1.0 & 
1.4 & 9.0 & 1,2$^a$ \nl
3C 147   & 0.55 & Q & 0.8  &  5.9  &  28.26  &   26.19  &  .013   &  .33 & 
24.0 & -- & 2   \nl
3C 268.3 & 0.37 & G & 1.4  &  8.6  &  27.13  &   24.17  &  .002   & .8 & 
-- & 0.2 & 1   \nl
3C 277.1 & 0.32 & Q & 1.7  &  9.7  &  26.86  &   25.02  &  .022   & .55 & 
12.0 & 31.0 & 4,1$^b$   \nl
3C 303.1 & 0.27 & G & 1.8  &  9.2  &  26.37  &  $<$24.1 & $<$.008 &  
$<$3.2 & -- & 28.0 & 1  \nl
3C 305.1 & 1.13 & G & 2.5  & 21.4  &  27.62  &  $<$25.3 & $<$.006 &  
$<$2.4 & 4.78 & - & 5    \nl
3C 343.1 & 0.75 & G & 0.4  &  3.2  &  27.78 &  $<$26.0 & $<$.03  & $<$10 & 
&  \nl
4C 12.50 & 0.12 & G & 0.08 &   .23 &  26.36  &   24.86  &  .21    & 8.4   
& -- & 48.0 & 1  \nl
\enddata 
\end{deluxetable}

\begin{deluxetable}{lccc} 
\footnotesize 
\tablecaption{CSS photometry} 
\tablenotetext{1}{Units of the line fluxes 10$^{-15}$ erg s$^{-1}$ 
cm$^{-2}$}
\tablenotetext{2}{Units of 10$^{-15}$ erg s$^{-1}$ 
cm$^{-2}$ \AA$^{-1}$ }
\tablewidth{0pt} 
\tablehead{ \colhead{Name} & \colhead{Total [O III] flux$^1$} 
& \colhead{Nuclear [O III] flux$^1$} & \colhead{Nuclear continuum 
flux$^2$} }
\startdata 
3C48    & Satur.  & Satur.   & 0.74 \nl
3C147   & 34 &  8  & 0.11 \nl 
3C277.1 & 38 & 13  & 0.19 \nl
3C303.1 & 35 &  7  & 0.013 \nl 
4C12.50 & 33 & 28  & 0.14 \nl
\enddata 
\end{deluxetable}

\end{document}